# Automation of MKID Simulations for Array Building with AEM (Automated Electromagnetic MKID Simulations)

Cáthal McAleer[1,2] · Oisin Creaner[2,3] · Colm Bracken[1,2] · Gerhard Ulbricht[1,2] · Mario De Lucia[2] · Jack Piercy[2] · Tom Ray[2]

## Abstract

Microwave Kinetic Inductance Detectors (MKIDs) are photon detectors comprised of superconducting LC resonators with unique resonant frequencies corresponding to their geometrical structure. As each pixel has its own geometry, electromagnetic simulations by hand of every pixel in a kilo-pixel array are impractical. Simulating fewer pixels and interpolating in between risks reduced pixel yield in arrays due to overlapping resonant frequencies. We introduce a new software called AEM (Automated Electromagnetic MKID simulations) that automates the construction and simulation of every simulated MKID pixel in an array according to specified resonant frequencies and a $Q_c$ range. We show automated designs to have an increased pixel yield (avoiding loses due to interpolation completely), increased accuracy in resonance frequency and $Q_c$ values when compared to interpolated structures. We also demonstrate a simulated trial of AEM for 100 MKIDs between 4 and 8 GHz to produce MKIDs with accuracies of ± 0.2 MHz with a runtime of 10 h 45 min.

**Keywords** MKID · AEM · Automation · Simulation

## 1 Introduction

MKIDs [1, 2] are superconducting LC resonant circuits that utilize the kinetic inductance effect to act as photon detectors for astronomical applications. Large MKID detector arrays with up to 20,000 pixels [3] have been achieved, and 2,000 individual pixels per feedline are standard [4]. Each pixel in a feedline has a unique geometry that defines its individual resonant frequency for frequency domain

✉ Cáthal McAleer
cathal.mcaleer.2018@mumail.ie

1   Department of Experimental Physics, Maynooth University, Maynooth, Co. Kildare, Ireland

2   Dublin Institute for Advanced Studies, 31 Fitzwilliam Street, Co. Dublin, Ireland

3   Dublin City University, Glasnevin, Co. Dublin, Ireland

multiplexing using a single feedline. For MKIDs with separate inductor and capacitor regions (also known as a lumped element KID [5]—LEKID), the capacitor is usually varied across pixels to achieve unique resonant frequencies.

In most MKID arrays, individual resonant frequencies $f_0$ of pixels on a feedline are designed to be equidistant in frequency space within a single octave. This restriction on the number of pixels per feedline originates from several factors. For optical to near-IR MKIDs, it is mainly the desired 1 μs time resolution, resulting in a 1 MHz frequency resolution of the Fast Fourier Transform (FFT) in the readout pipeline [6]. As such, resonators fabricated for optical to near-IR array purposes are usually spaced 2 MHz apart in frequency from one another to avoid "pixel clashing" where 2 resonators would end up in the same FFT bin, would therefore be indistinguishable and would need to be ignored.

As each individual MKID has a unique geometry and thus resonant frequency, simulations of these geometries are required during the detector array design phase. MKIDs are typically simulated using EM software such as Sonnet [7]. Sonnet allows MKIDs to be simulated as 2D planar superconducting structures using so-called Method of Moments analysis; however, designing the structure and fine-tuning for desired resonant frequency and coupling to the feedline are all done by hand. With a realistic approximate tweaking time of 30 min for a single MKID, simulating every pixel for a 2,000 pixels feedline becomes non-feasible, and as such, interpolation of resonant frequencies is usually used.

## 2 Interpolation & Automation of MKID Structures

Interpolation of MKID designs means that, for example, every 50th resonator is simulated and finely adjusted by hand, but the design of resonators in between (usually the length of one capacitor leg, see Fig. 4) is interpolated. This is a quick method to reduce simulation work required as it avoids many simulation hours but increases the risk of individual pixels clashing in frequency space (please see below). Several further effects like fabrication inaccuracies also lead to clashing pixels, and distinguishing between these has so far proved impractical. The best pixel yields to date are approximately up to 80% [8]. Significant improvements are possible with post-cooldown adjustments performed on the MKIDs such as trimming but have so far only been successfully demonstrated [9, 10] for far-IR and sub-mm MKIDs due to their significantly larger sizes, lower resonant frequencies and smaller arrays.

Interpolation methods currently demonstrated use higher-order polynomial fits to the length of a single capacitor leg as a function of resonant frequency [11]. This is, however, shown to either require many simulated resonators for accurate estimations of the remaining pixels or results in large deviations from designed resonant frequency.

It should also be noted that current methods of array interpolation typically focus on estimating resonant frequencies, and not on the coupling quality factor ($Q_c$) of the resonator. $Q_c$ dictates how well the resonator couples to the feedline and the shape of the resonance curve. $Q_c$ can become difficult to control for MKIDs of lower resonant frequencies and can be extracted from Sonnet as Sonnet assumes an infinity internal quality factor, and therefore, $Q_c$ is equal to the total quality factor in a Sonnet simulation.

Fully simulating every MKID pixel (for both resonant frequency and $Q_c$) would avoid any risk of the interpolation process to contribute to pixel losses. The automation of MKID simulations can significantly reduce the man hours required to simulate all pixels. Therefore, we introduce an automation software designed to interface with Sonnet and construct and simulate every MKID structure on a feedline for desired design parameters, thus avoiding the need for interpolation methods avoiding human bottlenecking; we call it Automated Electromagnetic MKID simulations (AEM).

AEM is a MATLAB script that analyses the simulation file Sonnet outputs and extracts a resonator's resonant frequency and quality factor. It afterwards adjusts the MKID pixel's capacitor and hands the updated geometry back to Sonnet. The interfacing between AEM and Sonnet uses the SonnetLab toolbox command set. The full AEM software package can be found and is freely accessible on our GitHub repository [12].

## 3 Method of Automation: AEM—Automated Electromagnetic MKID Simulations

We developed AEM as a software designed to interact with the EM simulation suite Sonnet to automate the construction and simulation of MKID pixels for specific, user-defined resonant frequencies and a range for acceptable $Q_c$ values. AEM requires an input of a general starting MKID design consisting of a feedline and surrounding ground plane with a boxed cavity for the MKID (see Fig. 1). The starting design also requires an MKID pixel without capacitor legs but with an empty "capacitor area" (i.e. defined polygons for the perimeter of the interdigitated capacitor (IDC)). The user provides AEM these specified geometrical dimensions, a list of resonant frequencies required and a desired $Q_c$ range for the MKIDs. All further design optimization and simulation is completely automated by AEM. The method of construction iterates between solving for resonance frequency and $Q_c$ of the structure, and all simulations

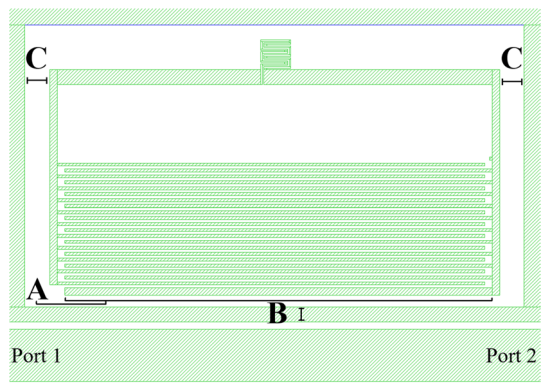

**Fig. 1** MKID structure simulated with Sonnet, $f_0 = 6490.6$ MHz. The labelled structures are optimized automatically by AEM and are: **A** coupling bar length and width, **B** width of the ground bridge, **C** distance to the ground plane

are performed by Sonnet's EM engine. As a simplification, AEM does not take neighbouring pixels into account at the moment, which would increase its accuracy and is planned in the future.

Starting with the smallest IDC, AEM optimizes for the largest desired resonant frequency first and proceeds with smaller frequencies. The length of the topmost leg of the IDC is first adjusted to find the user-defined resonant frequency ($f_0$) (if that is insufficient, another IDC leg is added). This is approximated using a "binary search" algorithm in which the software uses interval halving to search for two resonator structures with $f_0$ in between, where $f_0$ can't be achieved with better precision due to a step size defined in AEM for this first approximation. We usually use 10 μm here to reduce the number of required simulations. The solution with the higher frequency is chosen, and then, $Q_c$ can be solved by optimizing A, B and C (see Fig. 1). This will change the resonant frequency again; therefore, AEM in the next step checks resonant frequency which is still in between the two structures simulated before. If so, 10 simulations with 1 μm steps (our usual cell-size in Sonnet) for the topmost capacitor leg are performed to find the best approximation for $f_0$. If the resonator is no longer between the two structures simulated before, AEM goes one step back and does interval halving again. In the last step, AEM checks if $Q_c$ is still acceptable and repeats if necessary. This process steps through all resonant frequencies designated by the user starting from the previously "solved" structure. With our example step size of 1 μm, typically 15–20 simulations are required per pixel.

$Q_c$ is influenced by all changes to the MKID geometry and surrounding structures, and thus, multiple parameters can be varied to arrive at a structure with correct $f_0$ and $Q_c$. For AEM, the structures swept in order of operation are the coupling bar length and width A, the ground bridge width B and the distance to the ground plane C. The order in which to vary these structures was chosen due to their effects on $Q_c$ and space in array building. Variation in B produces the most significant effect on $Q_c$, with C showing the least effect. This can be explained as the MKID being predominately coupled to the feedline and ground plane, resulting in large effects on $Q_c$ by the distance and width of the ground bridge to the MKID. However, it was chosen to begin $Q_c$ parameterization with the coupling bar length due to being able to vary it in both length and width to a significant extent. In Fig. 2, we show for example the influence of the length and the width of the coupling bar on achievable $Q_c$ values. In Fig. 3, we demonstrate the sensitivity of $Q_c$ with variations in A and B.

Unfortunately, Sonnet is not absolutely stable and crashes from time to time. AEM therefore monitors Sonnet crashes by analysing the output file. We also check for non-physical result from Sonnet simulations by flagging feedline transmission values above one or unexpectedly low transmission values as faulty. In this case, AEM cleans the geometry file and repeats the simulation in a slightly adjusted frequency range until correct data are produced.

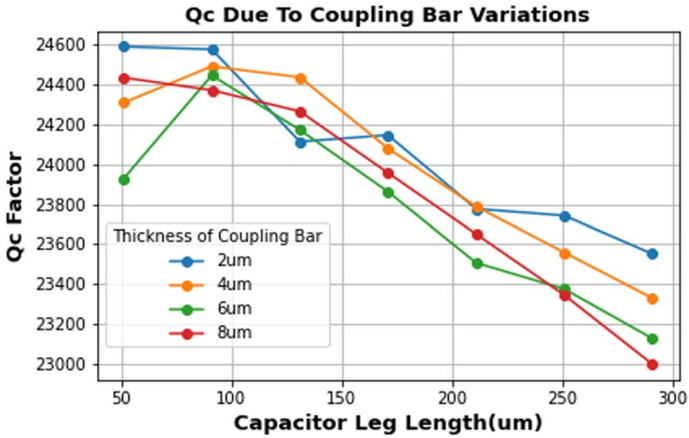

**Fig. 2** Effect on $Q_c$ by variation in coupling bar thickness and length. The noisy behaviour (for example seen in the 6 μm line) is an artefact of mainly the step size of the simulation result

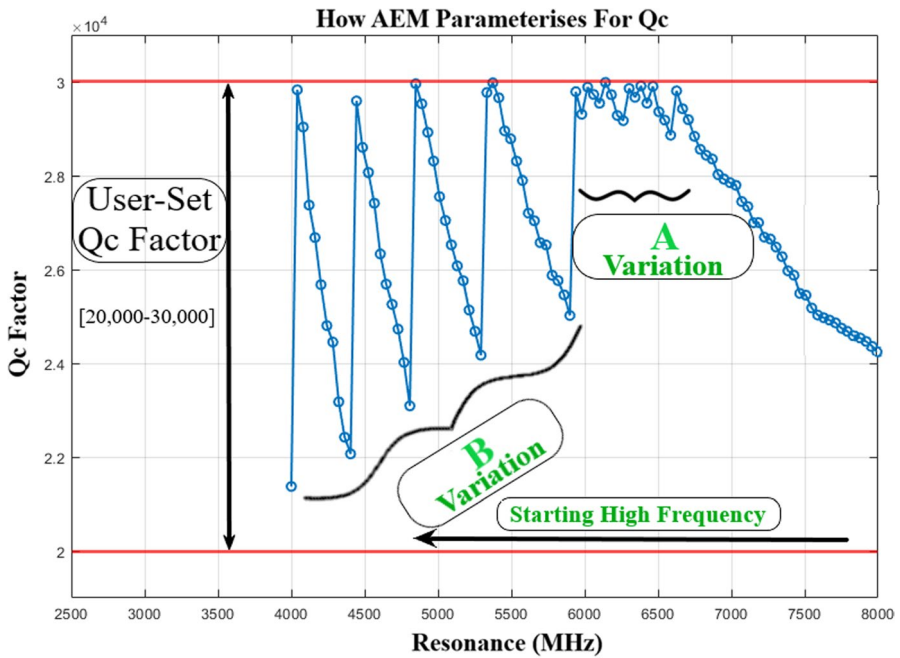

**Fig. 3** AEM method of solving for $Q_c$. 100 MKIDs are automated to be between $Q_c$ of 20,000 and 30,000. The coupling bar is initially varied in both thickness and length to solve for lower frequencies). For frequencies below 6 GHz in this example $Q_c$ becomes difficult to control with small changes in structure, and thus, ground bridge variations are preferred due to their stronger effect on $Q_c$

## 4 Results & Performance

To validate how much fully simulating an array could improve pixel yield over interpolation, we studied an example group of 57 resonators between 6501.2 and 6388.9 MHz with the length of their last capacitor fingers between 2 μm and 286 μm, respectively. For one data set, we simulated both edge resonators fully and linearly extrapolated the capacitor finger length for the 55 resonators in between for a desired 2 MHz spacing in frequency (Fig. 4, green points). According to this linear interpolation, the step size between capacitor lengths should be approximately 5 μm. The same group of 57 resonators was then constructed with AEM to automate finding the optimal capacitor lengths for each exact resonant frequency (Fig. 4, red points). We compare both methods by the distance in frequency space from the intended designed value (Fig. 4, left) and if pixels would fall into the same FFT bin before fabrication (Fig. 4, right). Resonators are classed as "clashing" if they have less than 1.5 MHz distance to their nearest neighbour as they then risk falling within the same 1 MHz FFT window. For this definition, we assume an additional 0.5 MHz movement of the resonances caused by fabrication inaccuracies and increased by the typical 0.2–0.3 MHz resonance width.

Figure 4 shows that the interpolated values lead to systematic deviations of above 6 MHz from the intended design, while the automated geometries only deviated by less than 0.3 MHz. Deviations in the automated values, specifically around the 6400 MHz and 6470 MHz range, are caused by the used cell size of 1 μm chosen in Sonnet. AEM is programmed to design a resonant structure as close to the desired resonant frequency as possible, and thus, the limiting factor in resonant frequency accuracy is purely down to the simulation's cell size, which has been chosen to match our available fabrication capabilities of 1 μm critical dimension lithography.

The important deviation in the interpolated and automated geometries can be seen much better when looking at clashing resonators that risk falling within the same 1.0 MHz FFT bin as these resonators would be lost in readout and hence would reduce the overall pixel yield. The pixel yield for the interpolated and automated groups in our example in Fig. 4 is 86% and 98%, respectively, showing a clear improvement with automated geometries by AEM. It should be noted that the one resonator lost with AEM sits exactly at 1.5 MHz to its nearest neighbour and may still be usable. Even though a single clashing resonator always makes two pixels unusable, we calculated the above yield with only one lost resonator per clash as our definition of clashing at 1.5 MHz distance in frequency is rather strict.

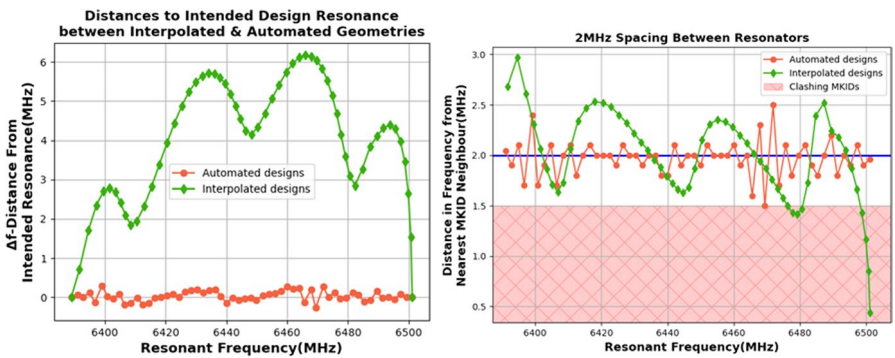

**Fig. 4** Resonators simulated and interpolated between 6501.2 and 6388.9 MHz by varying a single capacitor finger's length. Left demonstrates large deviations in resonant frequency by the interpolation method can be seen. The shape of the curve is likely caused by effects of the electric field of the capacitor as the minima occur across 1/4, ½ and 3/4 of the total length of the capacitor finger. Right shows all resonators that are lost due to "clashing" if they are less than 1.5 MHz in frequency space from neighbouring MKIDs

Our simulations of course do not take any fabrication inaccuracies into account. The typical pixel yield in MKID arrays of > 70% due to resonator collisions is generally attributed to fabrication defects, like deviations from design geometries or inhomogeneities in the superconducting layer. Figure 4 shows that interpolation could also contribute to this typical loss of pixels, dependant on for example, interpolation distance and pixel geometry. It is unfortunately not possible to make reliable predictions on the number of pixels lost due to fabrication inaccuracies, but AEM shows that resonant frequency interpolation can potentially add further loses, a risk that should be kept in mind.

For a better estimate on the required simulation times, we performed AEM simulations for a small-scale prototype with a 100 pixels array. 100 resonators would be time-consuming to construct by hand, and interpolation would as discussed reduce the expected pixel yield of the mask. To test AEM's array building performance, we went for 100 MKIDs with 5pH/sq for the superconducting film, designed to be equidistant (approx. 40 MHz) within the 4–8 GHz octave. The initial parameters given for this run are shown in Table 1.

The automation had finished the 100 MKIDs (some examples shown in Fig. 5) with a runtime of 10 h and 45 min and a total of 1457 simulations performed. All resonators lay between $Q_c$-20,000 and 30,000 (see Fig. 3) with a mean resonant frequency accuracy of $\pm 0.188$ MHz. This run was performed on a not especially strong PC using a 12 thread CPU and 16 GB of RAM. To date, Sonnet allows up to 64 threads [7] for calculations on a single machine, and thus, this simulation trial can be expeditated to be much faster on a more specialized computer.

The number of simulations performed is the result of two main causes:

(1) More sophisticated optimization method then binary search could allow to further decrease the number of simulations performed.
(2) Crashes & non-physical data produced within Sonnet occur with a rough rate of one in about 20 simulations. Further updates of the Sonnet lab toolbox by the manufacturer could offer further improvements.

Using the results above, 2,000 MKIDs simulated lying within the 4–8 GHz octave can be estimated to have a runtime of 215 h or roughly 9 days with similar accuracies for $f_0$ and $Q_c$. Assuming linear dependency on utilized CPU cores, this could likely be reduced to about 40 h on a modern CPU. Further improvements by providing more RAM are expected to be less significant.

**Table 1** Initial starting dimensions for the simulation test; details see text

| Capacitor leg thickness | Capacitor leg spacing | Initial coupling bar thickness | Number of MKIDs & resonances | Qc range |
| --- | --- | --- | --- | --- |
| 2 μm | 2 μm | 4 μm | 100 MKIDs, 4–8 GHz | 20,000–30,000 |

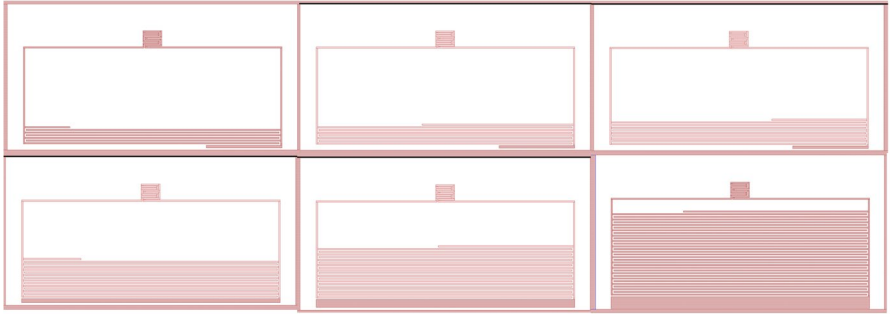

**Fig. 5** Resonators from the 100 pixels array automated by AEM (8000.9 MHz, 7313 MHz, 6666.4 MHz, 5454.6 MHz, 4767.7 MHz and 4000MHZ from top left to bottom right, respectively)

**Author contributions** C.McA wrote the main manuscript and prepared all figures. All authors reviewed the manuscript and provided comments and suggestions throughout the work.

**Funding** Open Access funding provided by the IReL Consortium.

## Declarations

**Competing interests** The authors declare no competing interests.

## References

1. P.K. Day et al., A broadband superconducting detector suitable for use in large arrays. Nature **425**, 817–821 (2003)
2. J. Zmuidzinas et al., Superconducting microresonators: physics and applications. Ann. Rev. Condens. Matter Phys. **3**, 169–214 (2012). https://doi.org/10.1146/annurev-conmatphys-020911-125022
3. A.B. Walter: MEC: The MKID Exoplanet camera for high speed focal plane control at the subaru telescope, PhD thesis, University of California Santa Barbara (2019)
4. S. Doyle: Lumped element kinetic inductance detectors, PhD thesis, University of California Santa Barbara (2008)
5. M.J. Strader: Digital readout for microwave kinetic inductance detectors and applications in high time resolution astronomy", PhD thesis, University of California Santa Barbara (2016)

6. N. Fruitwala: Readout and calibration of large format Optical/IR MKID arrays and applications to focal plane wavefront control, PhD thesis, University of California Santa Barbara (2021)
7. Sonnet Software Inc ,"Sonnet User's Guide", Release 16, May 2018, www.sonnetsoftware.com/support/manuals.asp
8. B. A. Mazin, et al: Astro2020 APC white paper optical and near-IR microwave kinetic inductance detectors (MKIDs) in the 2020s, Bulletin of the AAS 7(51), (2019)
9. X. Liu et al., Superconducting micro-resonator arrays with ideal frequency spacing. Appl. Phys. Lett. **111**, 252601 (2017). https://doi.org/10.1063/1.5016190
10. R. McGeehan et al., Low-temperature noise performance of superspec and other developments on the path to deployment. J. Low Temp. Phys. **193**, 1024–1032 (2018)
11. P. Szypryt et al., Large-format platinum silicide microwave kinetic inductance detectors for optical to near-IR astronomy. Opt. Express **25**, 25894–25909 (2017)
12. C. McAleer et al: AEM-Automated Electromagnetic MKID Simulations, version: 1.0.0, date-released: 2023-07-20, https://github.com/scathalmca/AEM/